# Are research contributions assigned differently under the two contributorship classification systems in PLoS ONE?

**Authors:**

**Kai Li (The University of Tennessee, Knoxville; kli16@utk.edu)**

**Chenwei Zhang (University of Hong Kong; chwzhang@hku.hk)**

**Vincent Larivière (University of Montréal; vincent.lariviere@umontreal.ca)**

**Abstract:**

Contributorship statements have been effective at recording granular author contributions in research articles and have been broadly used to understand how labor is divided across research teams. However, one major limitation in existing empirical studies is that two classification systems have been adopted, especially from its most important data source, journals published by the Public Library of Science (PLoS). This research aims to address this limitation by developing a mapping scheme between the two systems and using it to understand whether there are differences in the assignment of contribution by authors under the two systems. We use all research articles published in PLoS ONE between 2012 to 2020, divided into two five-year publication windows centered by the shift of the classification systems in 2016. Our results show that most tasks (except for writing- and resource-related tasks) are used similarly under the two systems. Moreover, notable differences between how researchers used the two systems are also examined and discussed. This research offers an important foundation for empirical research on division of labor in the future, by enabling a larger dataset that crosses both, and potentially other, classification systems.

## 1 Introduction

Our contemporary research system is built upon a greater level of collaboration between researchers with distinct expertise (Larivière et al., 2015; Wuchty et al., 2007). The larger number of collaborators in scientific research significantly challenged the traditional model of authorship that functions as the basis of the academic reputation economy (Cronin, 2001; Holcombe, 2019). Such challenges include the problematic practice of ghost and honorary authorship (Mowatt et al., 2002; Wislar et al., 2011) and the fact that multiple methods to determine the author order are being used in different research communities (Frandsen & Nicolaisen, 2010; Waltman, 2012). They further introduce noises and irregularities to the author order, creating more challenges to evaluate the importance of authors in a publication and in the whole research system.

In order to address the above difficulties in delineating authorship in the academic system, the concept of contributorship was first proposed in the late 1990s as an alternative approach to defining the relationship between authors and publications (Rennie et al., 1997). In this new model, the authors need to add a detailed description of the tasks undertaken by each author, instead of just listing all the authors after the publication title (Rennie et al., 1997). This can help researchers to receive more attention to what they did in the research and create a more transparent research ecosystem (Allen et al., 2014). The contributorship model was first adopted by biomedical journals soon after it was proposed (Smith, 1997), followed by some major publishers and journals, especially the Public Library of Science (PLoS) from the beginning of its journals in 2003 (Larivière et al., 2016).

As more journals moved to embrace this new model to replace and complement authorship, a problem in the infrastructure gradually emerged—the traditional five-task and idiosyncratic classification system was too coarse to capture the necessary details of author contributions. This is the original motivation for the International Workshop on Contributorship and Scholarly Attribution that was held in 2012, where a pilot project was established between various publishers, funders, and researchers to develop a cross-disciplinary taxonomy for contribution types (Hames, 2012). A new classification system composed of 14 roles was created from this project, which expands the traditional five-task taxonomy (Allen et al., 2014). This new classification system was further updated and added current name of CRediT, or Contributor Roles Taxonomy (Brand et al., 2015). It has also been heavily adopted by various publishers, especially by PLoS in 2016 to replace the traditional contributorship taxonomy (Atkins, 2016).

The accumulation of contributorship statements from journals have enabled a new research direction – investigating the division of labor using this novel data source supplied by authors themselves. A number of empirical studies have been conducted based on this type of data to investigate such questions like how research tasks are assigned to authors (Corrêa Jr et al., 2017; Larivière et al., 2016, 2020), the types of authorship in terms of the tasks one undertakes (Lu et al., 2020; Xu et al., 2022), the division of labor for researchers in different genders (Macaluso et al., 2016) and in different positions of the author list (Lu et al., 2022). These studies have effectively contributed to a deeper understanding of how scientific labor is divided within the social structure of research communities.

However, one notable limitation in these studies is that they have to rely on limited datasets, as not all journals have adopted those practices. First, nearly all of these studies used the contributorship statements from one or more PLoS journals—with a partial exception of Robinson-Garcia and colleagues (2020) who combined PLoS contributorship statements and author-level information from the Web of Science. Second, given the switch of contribution classification systems in PLoS journals, all of these research have to rely on a part of the data that is either based on the old or the new system. This limitation is also compounded by the fact no mapping scheme between the two systems have been created, that those tasks do not entirely overlap, and that a different classification scheme may lead to a different usage by researchers. All these factors have greatly limited the scope of conclusions we can draw from the emerging data source of contributorship statements.

To address the gap above in the literature, this research aims to understand how research contributions are assigned differently under the old and new contributorship classification systems adopted by PLoS journals. The comparisons are supported by our effort to develop a mapping scheme to map all tasks from the two classification systems into five categories that correspond to the five major tasks in the old system. In particular, we strive to understand the above broad question from the following perspectives:

RQ1: What share of articles use the old and new tasks?

RQ2: What share of authorship is assigned to the old and new tasks?

RQ3: Are old and new tasks assigned similarly across author order?

RQ4: Are old and new tasks undertaken by researchers in both genders similar?

This research represents the first project, to our best knowledge, to understand and contribute to the mapping between different contributorship classification systems, despite the growing body of empirical research using this data source to investigate the division of labor in research. We are hoping our research could inform future efforts to bridge the contributorship data based on different classification systems in PLoS journals, if not from other journals. Moreover, this research also aims to shed fresh light on the impacts of metadata standards on scientific works, and particularly how the implementation of the CRediT system shapes the assignment of scientific contributions to researchers.

## 2 Literature review

### 2.1 From Authorship to Contributorship

Several factors are considered by researchers when deciding authorship on a paper (Marušić et al., 2011). In a survey of 12 authorship guidelines from prominent organizations in a wide range of disciplines, such as the International Committee of Medical Journal Editors (ICMJE), the American Psychological Association, and the National Academy of Science, Osborne and Holland (2009) summarized that authorship should be granted to those who have made substantial contributions to the work and are responsible for the results. It has been a long time in the scientific practice that authorship should be “writing-mandatory” (Aliukonis et al., 2020). According to ICMJE, for example, “drafting the work or revising it critically for important intellectual content” is one of key criteria which qualifies a researcher to be an author of a manuscript (International Committee of Medical Journal Editors, 2018). Given the collaborative nature of contemporary science, researchers’ substantial contributions can be way beyond writing, which makes the concept of authorship not appropriate anymore (Holcombe, 2019). It has been suggested the writing mandate principle being removed from common authorship guidelines (McNutt et al., 2018).

As part of the bigger trend above, the contributorship model was first proposed by Rennie and colleagues, aiming to reflect the actual collaborations in scientific practice, where instead of simply listing the name of authors, a systematic description of who did what is suggested (Rennie et al., 1997). The contributorship model has been adopted by a few biomedical journals since the late 1990s (Smith, 1997). Under this model, all of those who make substantial contributions will be listed in the contribution list and each co-author’s number and types of contributions will be explicitly indicated (Sauermann & Haeussler, 2017). Contribution statements can bring various benefits, such as clarifying the roles of methodological innovators whose contributions are not apparent in a conventional authorship, and promoting data sharing by identifying the providers (Allen et al., 2014). It also aids the decision-making of different stakeholders, including research institutions, funding agencies, and scientific societies (McNutt et al., 2018).

According to Sauermann and Haeussler (2017), 11 out of 15 journals with highest impact factors in the category “multidisciplinary sciences" in the year of 2014 require the team to disclose each author’s contribution. However, only three journals, *Science*, *PLOS ONE*, and *Peer J*, have offered a standardized template for contribution disclosure, whereas other journals only ask for open-ended statements. A lack of standardized disclosure may hinder the evaluation and comparison of authors’ efforts across journals (Allen et al., 2014; Marušić et al., 2006).

### 2.2 Contributorship Classification Systems

A series of idiosyncratic classification systems for contributorship has been introduced by various journals over the last few decades to recognize each individual’s efforts and responsibilities for specific research tasks. These systems vary in the number of contributions characterized, the organization of different contributions, and the accuracy of coverage (Larivière et al., 2020). Some systems include hierarchical items, which makes a certain contribution a prerequisite for the others; and in other cases, such as The Journal of the American Medical Association (JAMA), there is a structured checklist for authors to select and quantify the number of different contributions made by each team member (Fontanarosa et al., 2017).

PLOS journals have been using contribution statements to describe tasks undertaken by each author since 2003 (Larivière et al., 2016). It initially established a five-role contributor taxonomy, where the five major tasks were identified for authors in those contribution statements, including “Analyzed the data,” “Conceived and designed the experiments,” “Contributed reagents/materials/analysis tools,” “Performed the experiments,” and “Wrote the paper” (Larivière et al., 2016; Lu et al., 2020). Some other minor tasks were recognized as well, such as “Revise Paper,” “Approve Paper,” “Supervised the Research,” and “Funding.” every author of a manuscript should contribute to at least one of the five major roles, but not necessarily the other minor roles.

CRediT (Contributor Roles Taxonomy) taxonomy was initiated in a collaborative workshop led by Harvard University and the Wellcome Trust in 2012 (Allen et al., 2014) to address limitations in the landscape of idiosyncratic classification systems of research contribution. It was finally established in 2015, with a standardized set of 14 research contributions. It is “the best currently available method for embedding authors’ contributions in journal metadata” (McNutt et al., 2018, p. 2558). The adoption of CRediT taxonomy helps mitigate the confusion on author order across different disciplines and cultures (Sauermann & Haeussler, 2017).

A growing number of journals and publishers have implemented CRediT, including eLife, Cell, F1000, PLOS, Elsevier, Springer, and BMJ. More than 120 journals had adopted this new taxonomy by early 2019 (Allen et al., 2019); while the adoption number by a single publisher, Elsevier, already reached 1,200 journals at the end of the same year (Elsevier, 2019). PLOS journals started to adopt CRediT in 2016 to replace its old 5-role contributorship system.

### 2.3 Contribution Statement for Scientific Collaboration Studies

Over the recent years, there has been a growing number of papers research analyzing contribution statements from research articles. The foci have been put on the following issues: the variability of contribution by authors, the types of authorship reflected by the tasks undertaken, and the composition and division of labor across teams, such as among authors in different positions, authors with different seniority, and authors with different genders. From the lens of authors’ contribution statements, it is clearly seen that different authors are performing different tasks in a collaborative team (Lu et al., 2020). The division of labor increases with the team size (Haeussler & Sauermann, 2020).

Larivière et al. (2016) analyzed 87,002 PLOS journal articles’ contributorship statements under its old taxonomy in which five major contributions were identified. By comparing the labor distribution across disciplines, authors’ order, and seniority, the authors found that the division is more extensive in medical research than natural sciences, and drafting and editing of the manuscript is the most common task among authors in most domains except for medicine. They also demonstrated a U-shaped relationship between extent of contribution and author order–first authors generally took more contribution roles, followed by last authors, while authors in the middle positions took the least proportion. Similar results have also been found by Sauermann and Haeussler (2017) in an analysis of the author contribution statements of all articles published in *PLOS ONE* between 2007 and 2011. They found the number of contributions taken per author decreases with the team size; the relationship between author order and contribution statements were not always aligned–more errors would occur when using the author order to infer the first authors’ contributions in larger teams. Larivière et al. (2020), in an investigation of the contribution statements of all *PLOS* articles published between June 15, 2017 and December 31, 2018, which adopted the new CRediT contributor taxonomy, also confirmed an inverse relationship between the tasks contributed by first authors and last authors, while the middle authors have contributed the least. By analyzing the major five contributions from the old *PLOS* taxonomy, Corrêa et al. (2017) identified three distinct patterns of contributions in articles collaborated by many authors. The contribution to tasks such as experiments performance made by authors diminishes with their rankings; the contribution to some tasks such as data collection increases from first to second-to-last authors; the contribution of some other tasks such as data analysis and manuscript writing by authors is symmetric, where the first and last authors contribute the most while the intermediary authors contribute the least.

By checking the relationship between the contribution made by authors and their seniority, it was found technical tasks are more likely to be undertaken by young authors while the conceptual contributions are more associated with senior scholars (Larivière et al., 2016). Macaluso et al. (2016) examined the five major contributions by authors of 85,260 articles published in seven *PLOS* journals between 2008 and 2013 and found that women were more likely to contribute physical labor, such as experiment performance, while male authors were more likely to be associated with conceptual labor, such as conceiving and designing experiments and writing the article. Such findings have been reinforced by Larivière et al. (2020). Males have also conducted more seniority associated tasks, such as funding acquisition and supervision. In addition, the authors found that women were more likely to write the original draft, while the males were more likely to be involved in the review and edit of the manuscript. From the more fine-grained contribution statements, the authors were able to correct the skewness in Macaluso et al. (2016)'s finding, which may be due to the ubiquity of the "review" portion of writing.

The comparison between the *PLOS* journal articles' contribution statements adopting the old *PLOS* contributorship taxonomy and those in the new CRediT has demonstrated some noticeable differences. Larivière et al. (2020) found that contributing to the original draft has been taken by a narrower proportion of authors under the CRediT taxonomy compared to those "writing the paper" under the old taxonomy (Larivière et al., 2016). The adoption of CRediT taxonomy enables the recognition and reward of specific contributor roles, such as data and software engineers (Ding et al., 2021). There is a growing number of research investigating the PLOS journal articles' contribution statements adopting the new taxonomy. de Souza et al. (2022) compared the publications from three mainstream medical journals, *Annals of Internal Medicine*, *Journal of the American Medical Association*, and *PLOS Medicine*, seeking to group different contribution categories for standardization cross journals.

However, one big limitation exists in the current analysis of contribution statements–the lack of integration between different contribution taxonomies and the consequent limited investigation scope that current analysis is usually restricted in either very few journals or within a short period. To address it, in this research we aim to integrate the old and new *PLOS* contributorship taxonomies by mapping different contribution tasks into standardized categories.

## 3 Methods

### 3.1 Dataset

This analysis is built upon all research articles published in the journal of *PLoS ONE*. Metadata of publications were collected from the Web of Science and the full-text publications (with paratext) were collected from the publisher's API being implemented into the *rplos* package in the R programming language (Chamberlain et al., 2021).

We specifically selected all publications from 2012 to the end of 2020, i.e., five years of publications before and after the switch of the classification system by the journal. A total of 198,465 research articles from the journal were included in our analytical sample, including 122,170 and 76,295 publications using the old and new systems, respectively. As shown in Figure 1, the adoption of the CRediT system by the Public Library of Science had an immediate impact on how the researchers choose from the two systems. The transition between the two systems happened in the year of 2016, with only 55 publications in 2017 still using the old system. To reduce potential noises, these 55 publications are removed from our analysis.

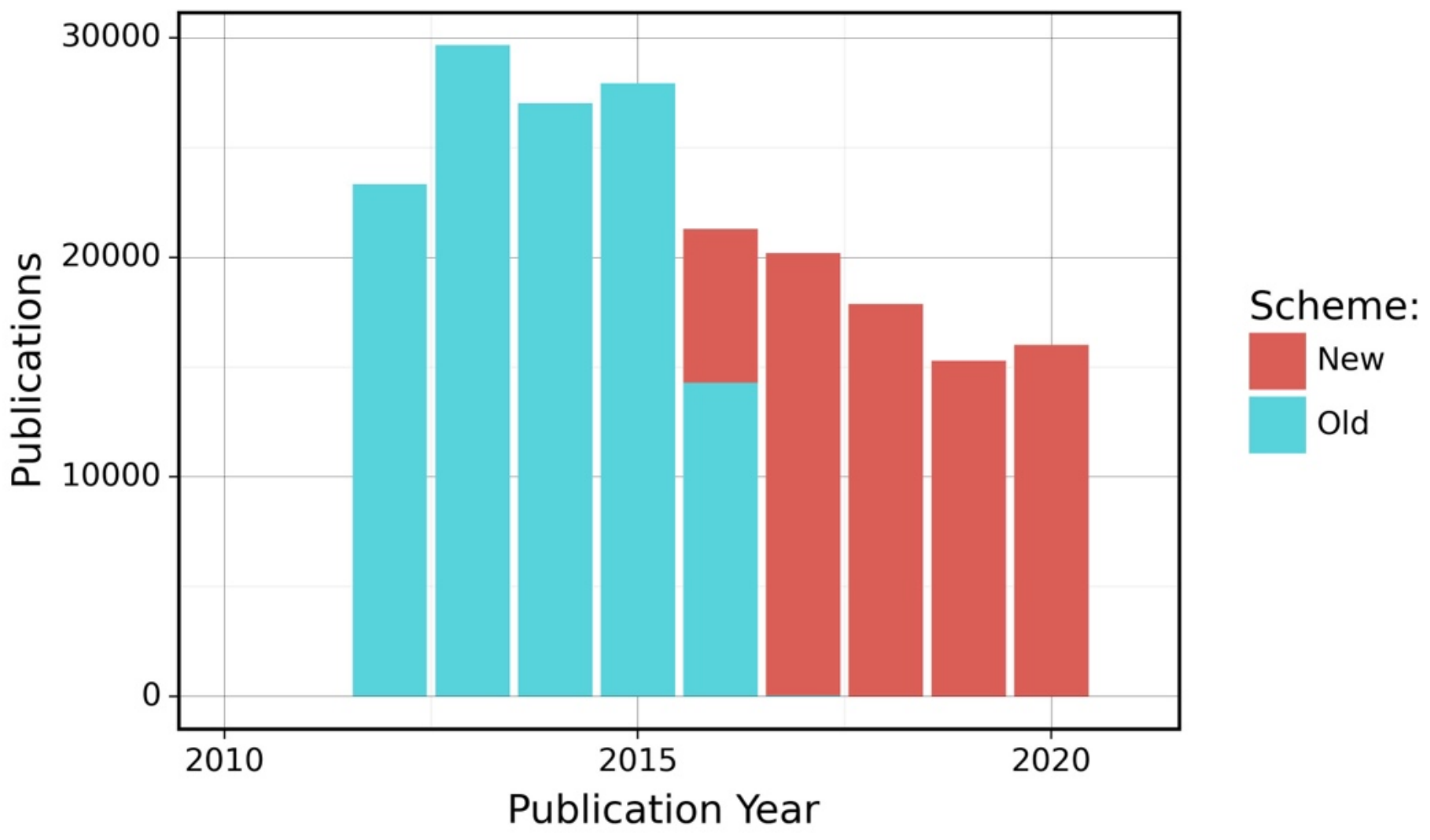

**Figure 1: Temporal distribution of the analytical sample**

### 3.2 Contribution extraction

We extracted contributorship statements from the downloaded XML full-text publications. The journal requires all publications to have a contribution statement that specifies what tasks were performed by which author of the paper (Robinson-Garcia et al., 2020). As described above, two different classification systems have been used by the journal: the old five-task system from the beginning of the journal in 2003 and the new 14-task CRediT system from 2016 (Atkins, 2016). All tasks in the two classification systems are listed in Table 2 in section 4.1.

Beyond their different tasks, another notable difference between the two systems is that the statements based on them have different formats in the full-text XML files. Under the old system, a statement is just a text string including all the author names (usually the acronyms of the names) and one's corresponding tasks, such as:

"SIC and GSB conceived and designed the experiments. SIC, CDVR, CC, AR, and SK performed the experiments. SIC, CDVR, CC, SK, and FM analyzed the data. YC and AR contributed reagents/materials/analysis tools. SIC and GSB wrote the paper."

Under the new system, all the statements are separated by tasks. However, before the year of 2019, most of the statements still used name acronyms to represent authors, such as: "Wrote the paper: EL HM". And only after that year, the majority of the statements move to a fully structured format to record the contribution of all authors, i.e., by using the order of authors in the author list.

For each format, we designed a distinct pipeline to map all the tasks to their corresponding authors. Specifically, we considered different ways in which author names are abbreviated. Moreover, we only focused on the five major tasks in the old system (and removed author-task pairs based on the minor tasks), because (1) all these minor tasks are used by relatively few researchers (pairs beyond the five major tasks only contribute to 3.2% of all author-task pairs using the old classification system) and (2) many of these tasks are difficult to be mapped to standardized categories (see section 3.5).

For all publications in our sample, we found 4,216,584 unique author-task pairs, including 2.19 million pairs based on the new classification system and 2.03 million based on the old system.

### 3.3 Normalized byline order

We used a standard method (Corrêa Jr et al., 2017) to calculate byline order of all authors according to their positions in the author list, to understand how the author order is related to tasks in both classification systems. Each author in a paper is represented by a ratio between one's position in the author list (from 1 to n) and the total number of authors (n). For example, in a two-author papers, the first author is represented as 0.5 and the second author is represented as 1.

In most of the analyses involving the byline order, we group publications according to the number of authors, in order to more accurately represent the relationship between tasks and byline order. We mostly only considered publications with up to 12 authors, which cover 183,465 publications, or 92.4% of our analytical sample.

### 3.4 Gender identification

To evaluate how tasks have been assigned under both systems from the perspective of the gender of researchers, we used the algorithm developed in Larivière et al. (2013) and Sugimoto and Larivière (2023) to automatically assign the gender of authors. The algorithm uses given names and family names of authors, and assigns them a probability of being men or women. The algorithm takes into account countries, as some names are associated with a different gender in different countries. For all 1,418,078 publication-author pairs (*authorship*) from our analytical sample, 1,251,947 of them (88.3%) are assigned a binary gender. To reduce noises connected to unisex names, only the latter sample of names is used in our analyses focusing on gender.

Among all the publication-author pairs with a binary gender, there are 470,211 pairs assigned to women (37.6%) and 781,761 pairs to men (62.4%).

### 3.5 Mapping of the two systems

We mapped the two classification systems in order to compare how they are used before and after the transition of the system. Table 1 summarizes the mapping scheme we developed based on our best interpretation of the definitions of all tasks in the two systems, where both types of tasks are mapped to the *categories* that corresponds to the old tasks.

**Table 1: Mapping between the two classification systems**

| Task category | Old tasks | New tasks |
|---|---|---|
| Conceive and designed | Conceive and designed the experiments | Conceptualization |
| Perform | Perform the experiments | Investigation |
| Analyze | Analyze the data | Formal analysis; Validation; Visualization |
| Contribute | Contribute reagents/materials/tools | Data curation; Funding acquisition; Software; Resources |
| Write | Write the article | Writing original; Writing review |

In particular, we should note the following observations made during the creation of the mapping scheme. First, there are a few tasks in the CRediT system that do not fit into the five tasks in the old system, such as *Funding acquisition*, *Project administration*, and *Supervision*. This corresponds to the CRediT system's goal to highlight researchers who may be shadowed by how labor is divided in the traditional research reward system (Holcombe, 2019). Second, we fully acknowledge that there are uncertainties in the boundaries between tasks in both the old and new systems, which makes it challenging to map between them. One notable example is the relationship between *Data curation* in the new system and *Contribute reagents/materials/tools* in the old, which may depend on how data is defined in specific research contexts. In this research, we chose to include all new tasks that may fit into the *Contribute* category. However, this decision itself is further examined and validated in this research to evaluate the validity of the mapping system and understand how the tasks are used differently before and after the adoption of the CRediT system.

### 3.6 Descriptive analysis

Results show that the number of authors per paper remains stable from the beginning of the journal until 2019, as shown in Figure 2. As a result, we can assume that the new contributorship policy has no direct impact on the size of research teams, despite the counterargument that the expansion of forms of labor contributed to the increasing number of collaborators in publications (Larivière et al., 2020; Shapin, 1989).

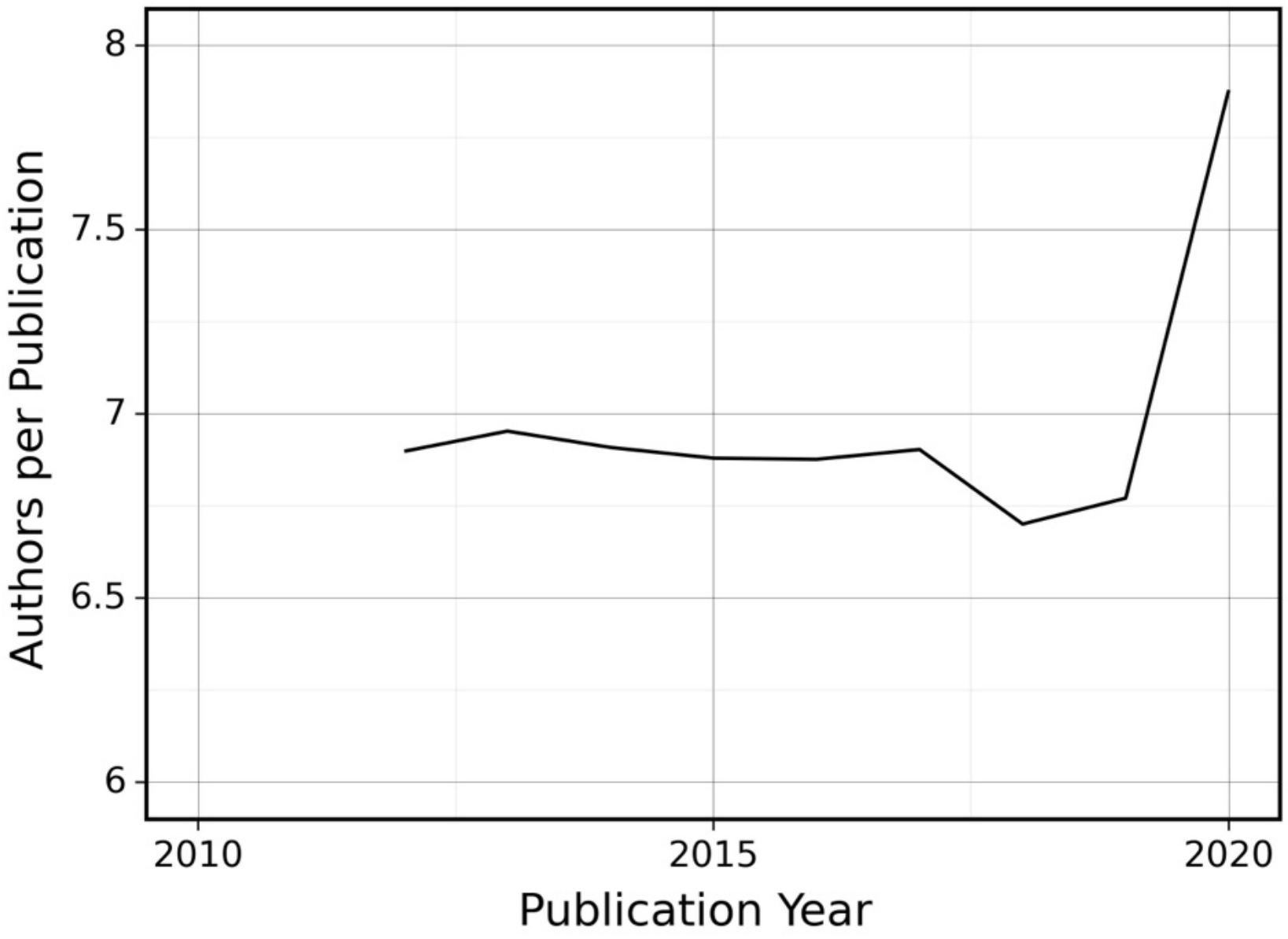

**Figure 2: Mean number of authors over time in our sample**

The above pattern also echoes the finding that the the composition of papers with different numbers of authors remain quite stable during the publication window we examined. As shown in Figure 3, the shares of papers using the two systems with up to 12 authors (the facet in the graph) are mostly identical and stable before and after the year of 2016.

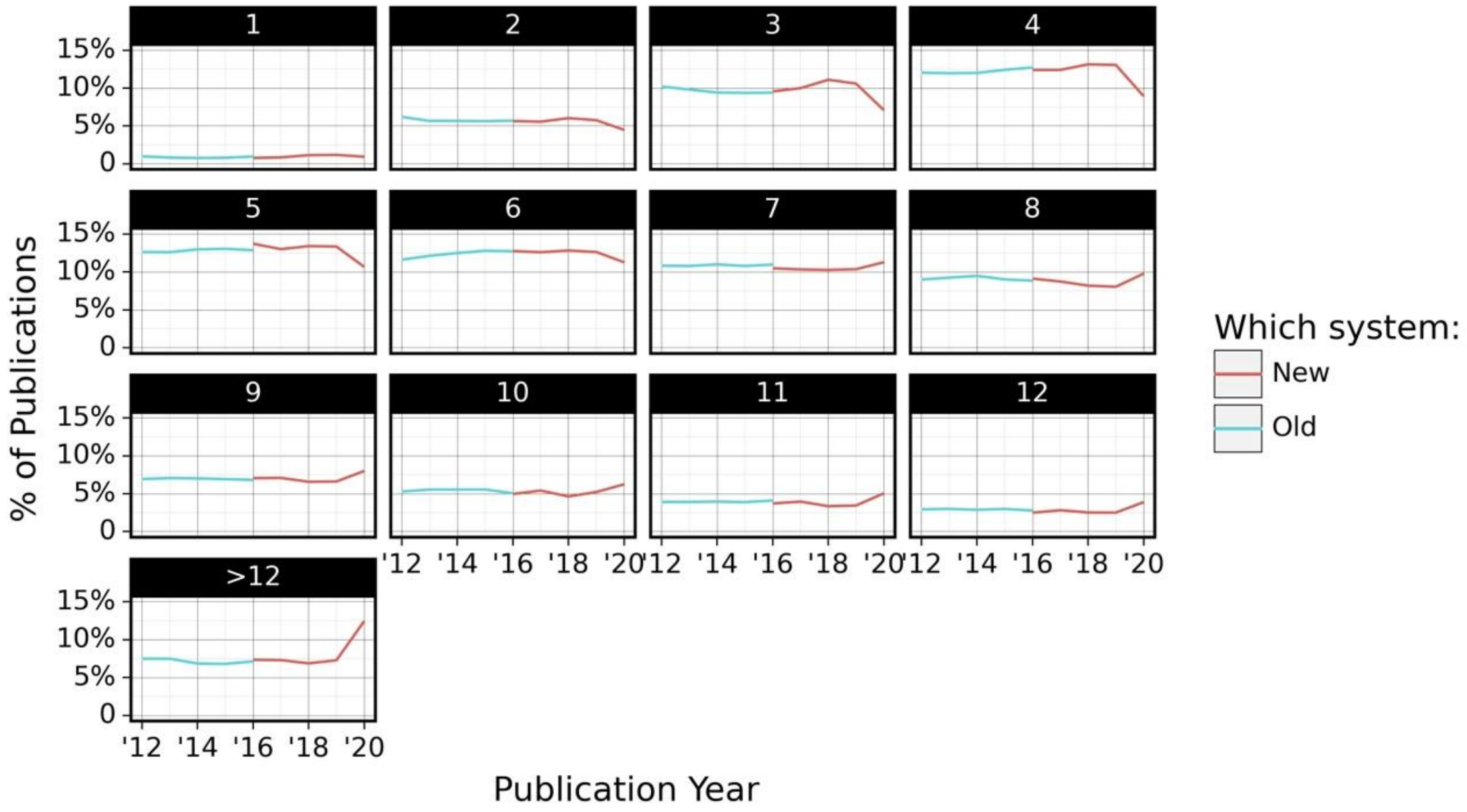

**Figure 3: Share of publications with different team sizes over publication years**

## 4 Results

### 4.1 How are categories used in all publications?

Table 2 summarizes how many papers in our analytical sample used all tasks included in both classification systems. The majority of old tasks are used in over 90% of all corresponding publications. Some of the new tasks are also heavily used, despite the much less even pattern among all new tasks.

**Table 2: Use of old and new tasks in our analytical sample**

| Scheme | Task | Publications | Share of Publications |
|---|---|---|---|
| Old | Write | 119,543 | 97.8% |
| | Analyze | 118,573 | 97.1% |
| | Conceive and design | 117,536 | 96.2% |
| | Perform | 113,833 | 93.2% |
| | Contribute | 92,563 | 75.8% |
| New | Writing original | 74,544 | 97.7% |
| | Writing review | 73,040 | 95.7% |
| | Conceptualization | 72,421 | 94.9% |
| | Formal analysis | 69,512 | 91.1% |
| | Methodology | 68,476 | 89.8% |
| | Investigation | 64,706 | 84.8% |
| | Supervision | 62,988 | 82.6% |
| | Data curation | 59,640 | 78.2% |
| | Project admin | 53,080 | 69.6% |
| | Funding acquisition | 50,644 | 66.4% |
| | Resources | 44,379 | 58.2% |
| | Validation | 43,661 | 57.2% |
| | Visualization | 42,107 | 55.2% |
| | Software | 30,176 | 39.6% |

Moreover, we also examined the above patterns over time. The results are shown in Figures 4 and 5 for each classification, respectively. Figure 4 shows that most of the old tasks were used in a very consistent way over time, with a notable exception of *Contribute* that shows a mild increase (and its increase is about 30% of publications from 2006 to 2016). As compared, in Figure 5, most of the top tasks are also used very consistently on the level of publications. However, some tasks, such as *Project administration* and *Funding acquisition*, show a decreasing trend from 2016 to 2020. The rise of old tasks in the data could be attributed to the fact that old tasks went through a process of normalization, where many of the minor tasks are standardized into this tasks over time.

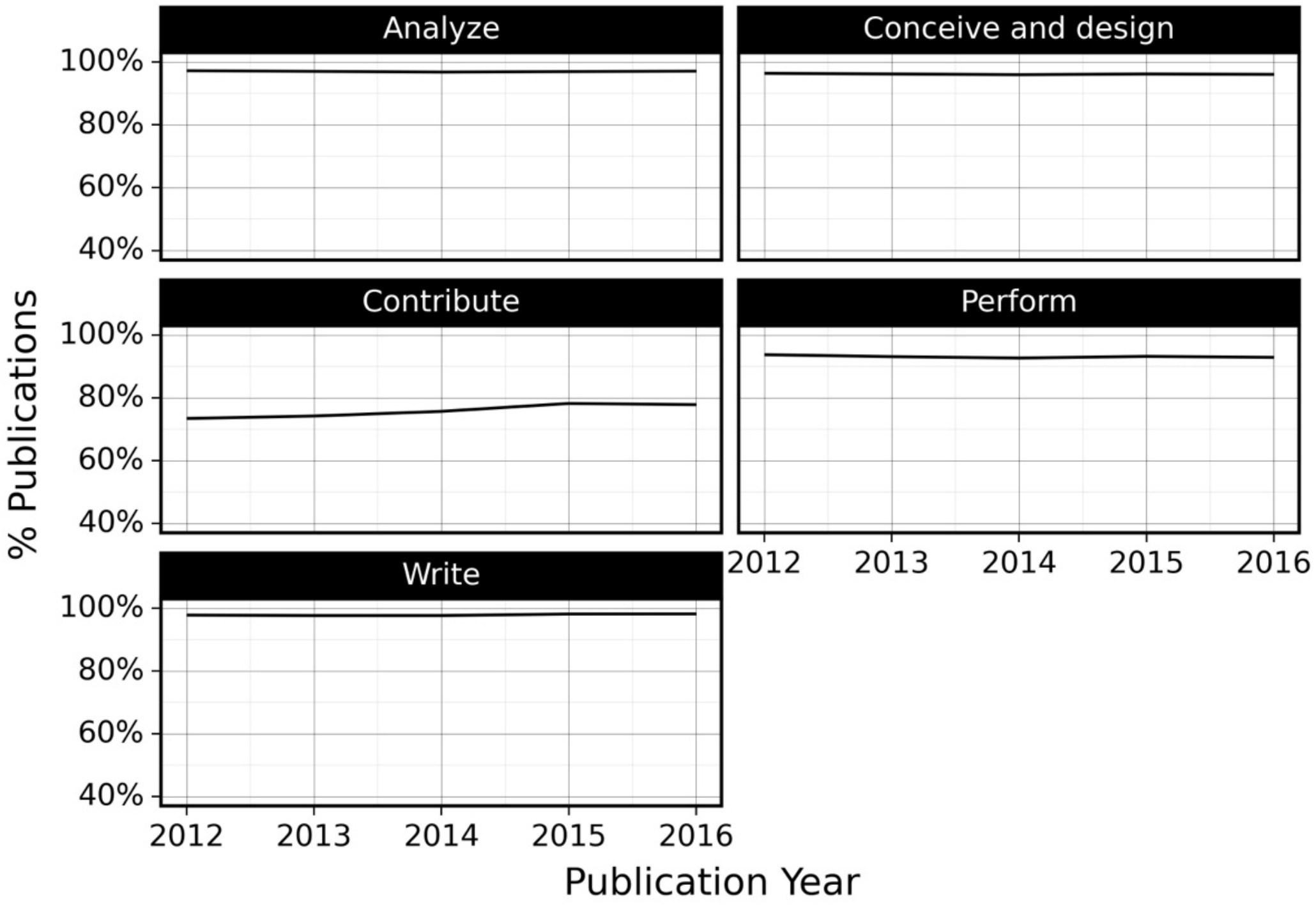

**Figure 4: Usage of old tasks over time**

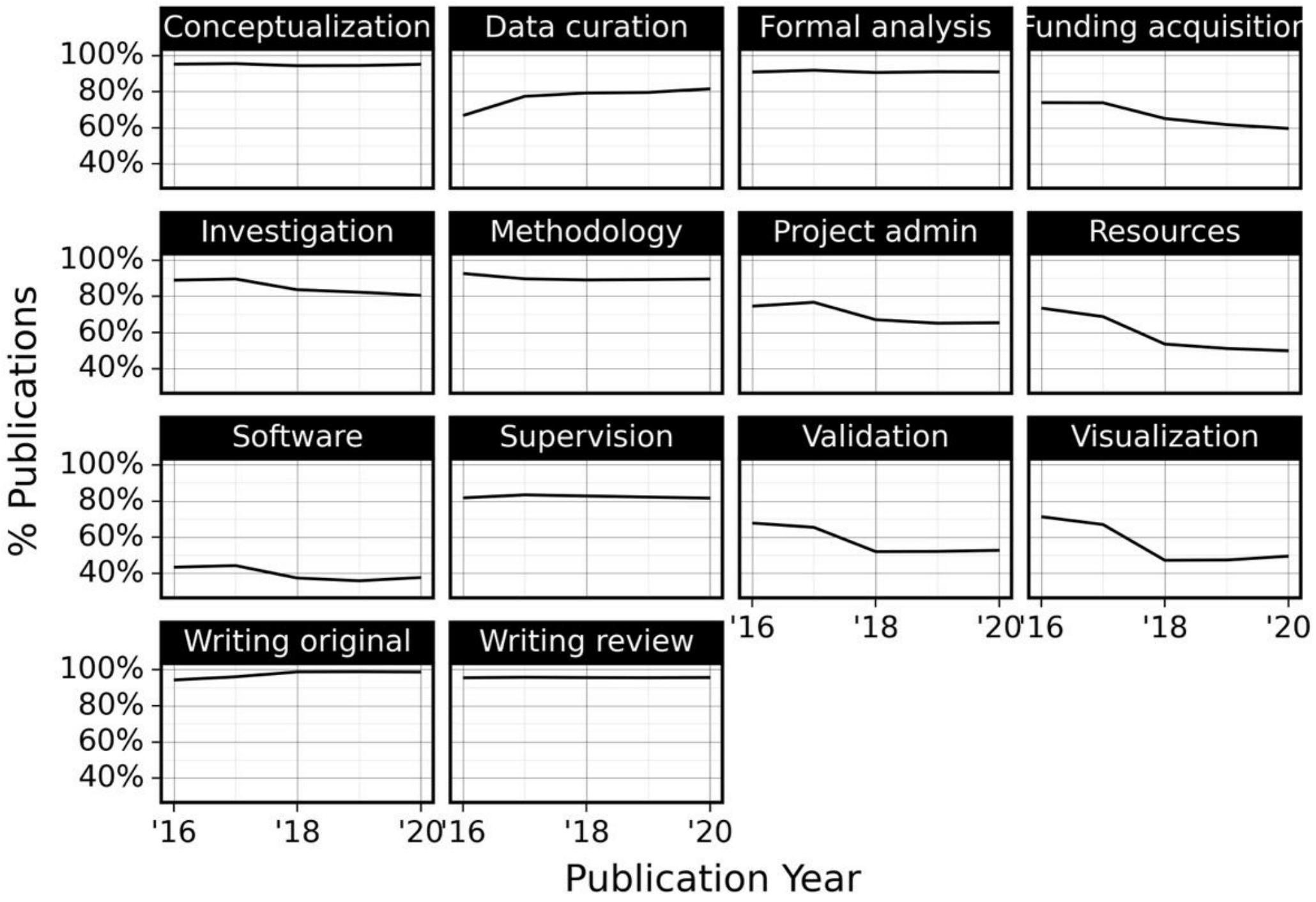

**Figure 5: Usage of new tasks over time**

Using our mapping scheme, we find that three of the five corresponding categories from the two systems are used similarly in the year of 2016 and within the 5-year windows of using the two individual systems (i.e., 2012-2016 and 2016-2020), as shown in Figure 6. However, *Contribute* and *Perform* have notable gaps between the two classification systems. This indicates that the new tasks mapped to *Contribute* may exceed its original scope and there is an opposite reason for *Perform*.

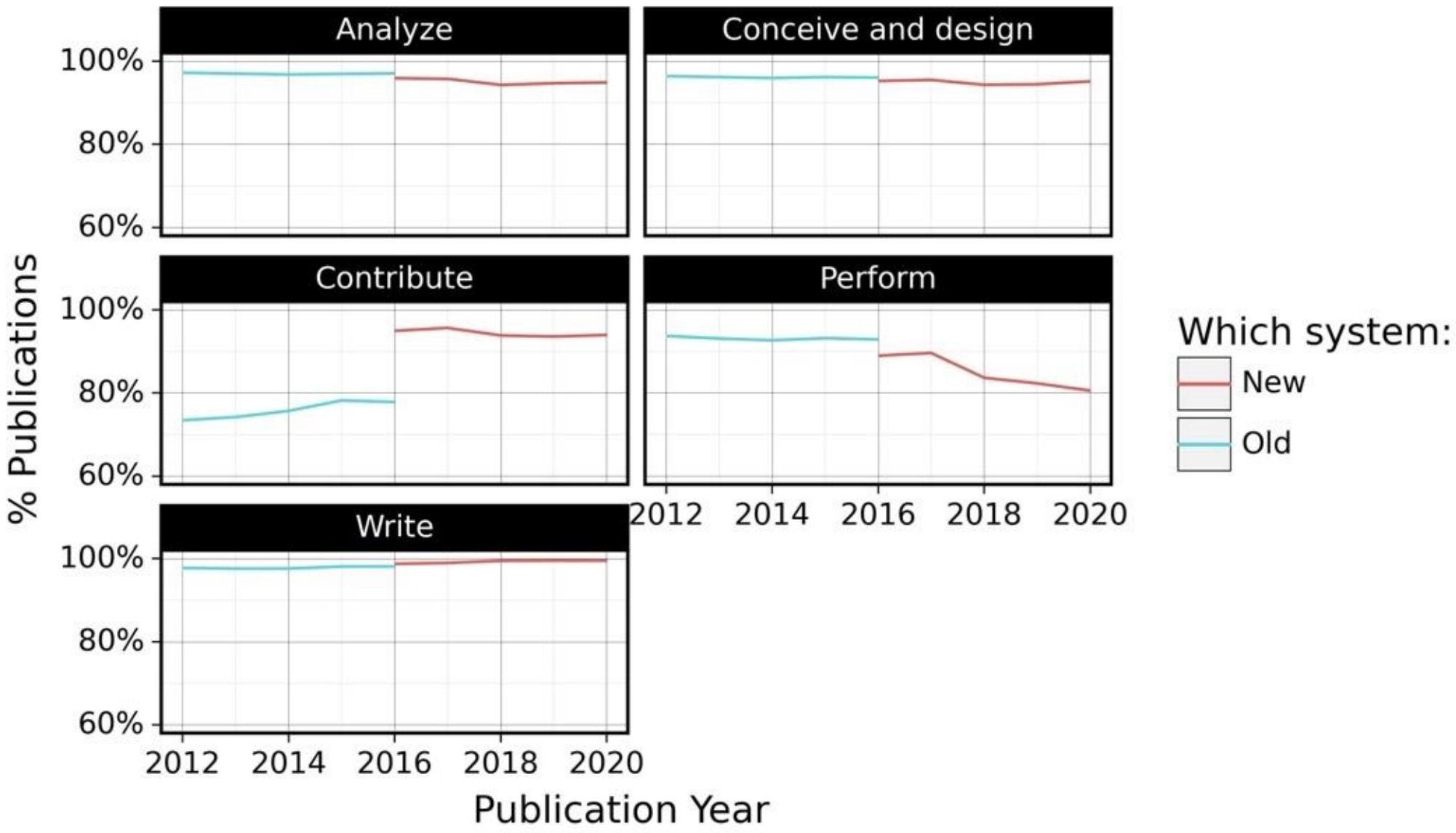

**Figure 6: Comparison of how the two systems are used in all publications over time**

## 4.2 Authorship in the two classification systems

By using the mapping scheme above, we calculated the share of authors performing each category of tasks in papers with various numbers of collaborators. In the graph below, the facets represent the five categories, which all old and new tasks are being mapped to. The graph shows that most of the tasks in each category follow a similar pattern as the number of collaborators increases, especially for the two categories where there is a one-on-one relationship between the two systems, i.e., *Conceive and design* and *Perform*. However, *Writing review* forms as a major outlier in the category of *Writing*: it is performed by a larger number of authors than in the old system, particularly in papers with more authors.

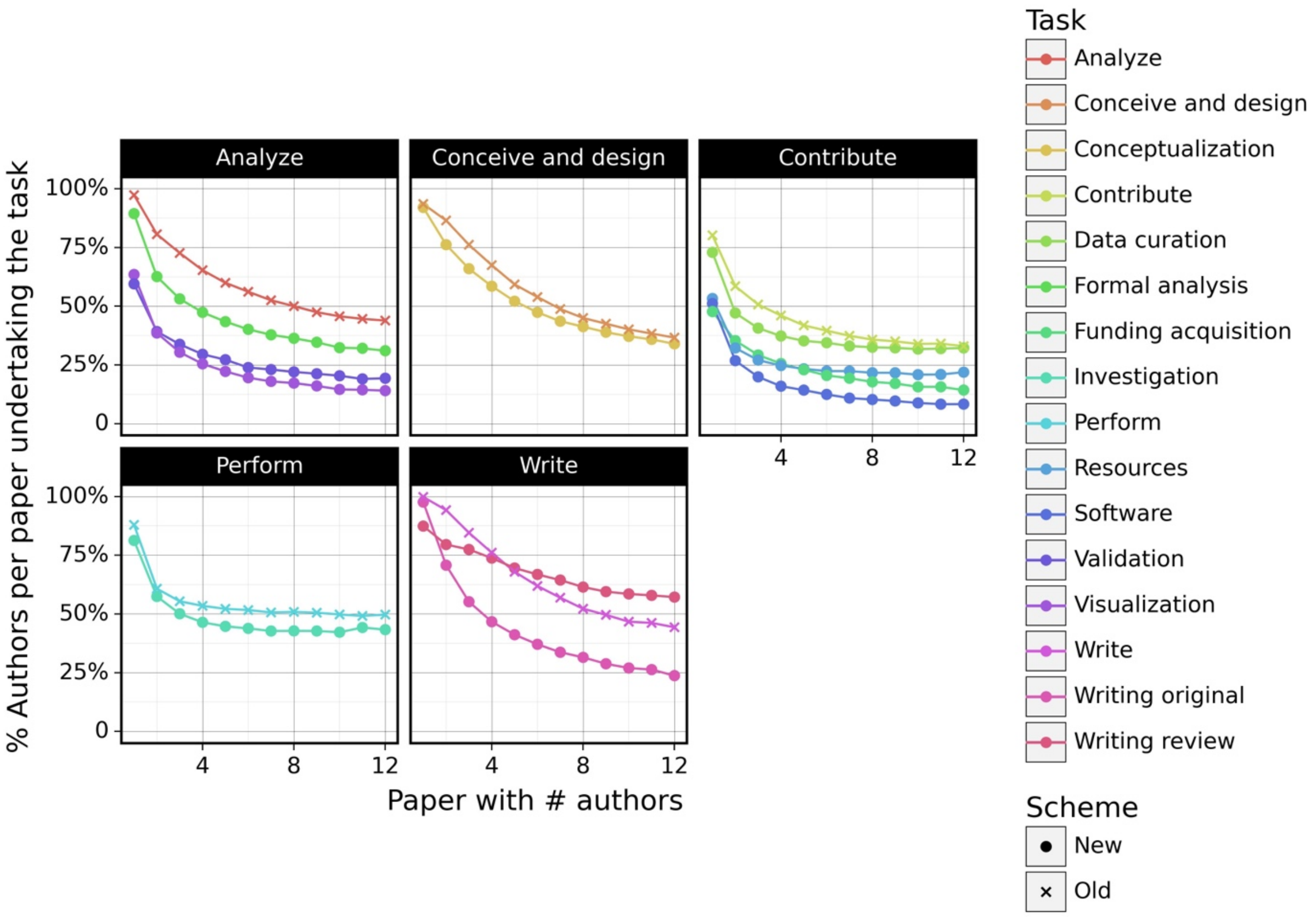

**Figure 7: Share of authors assigned to categories in the two systems in publications with different team sizes**

When the old and new tasks in the same category are combined together, the two systems tend to have very similar patterns across the five major categories, as shown in Figure 8. What is particularly interesting is the fact that in most of the cases, the old tasks are performed by slightly more authors than the new ones, except for the two outlying categories. It should also be noted that the pattern remains solid when we select smaller publication windows (i.e., three and four years) for both classification systems.

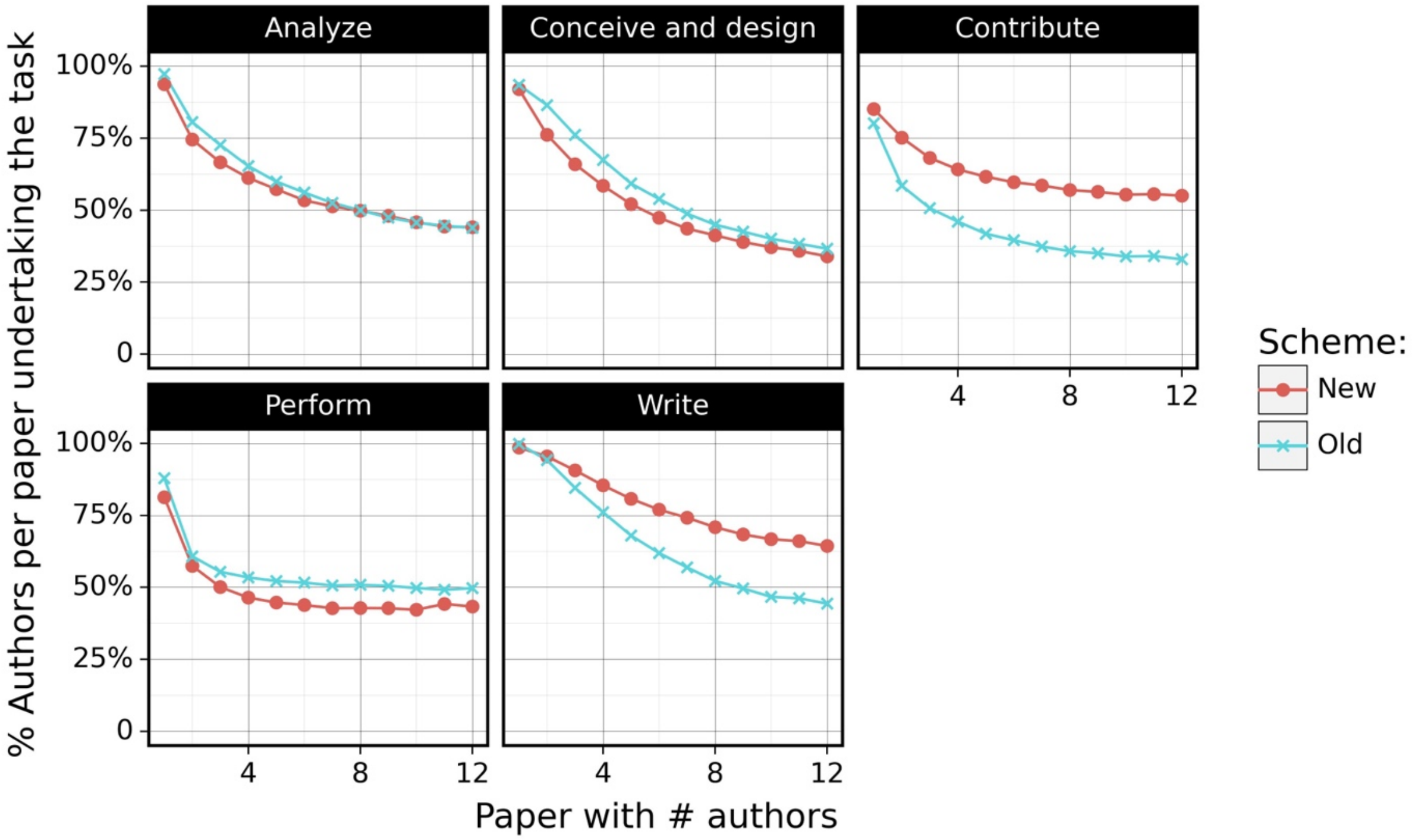

**Figure 8: Share of authors assigned to mapped categories in the two systems in publications with different team sizes**

### 4.3 Tasks and byline order

We further examined how authors in different positions of the byline undertake different tasks. Figures 9-11 show three examples from the five task categories. Figure 9 focuses on the *Analyze* category, which displays how the share of authors undertaking tasks under the two schemes (color) varies by the byline order (x-axis), in papers with various team sizes (facet). *Analyze* is an example where both schemes are being undertaken very similarly over the byline order. Combined with other evidence above, we can argue that the old and new tasks related to *Analyze* can be safely mapped. The same identical pattern between the old and new system can also be observed in the category of *Conceive and design*, as shown in Figure A2 in the appendix.

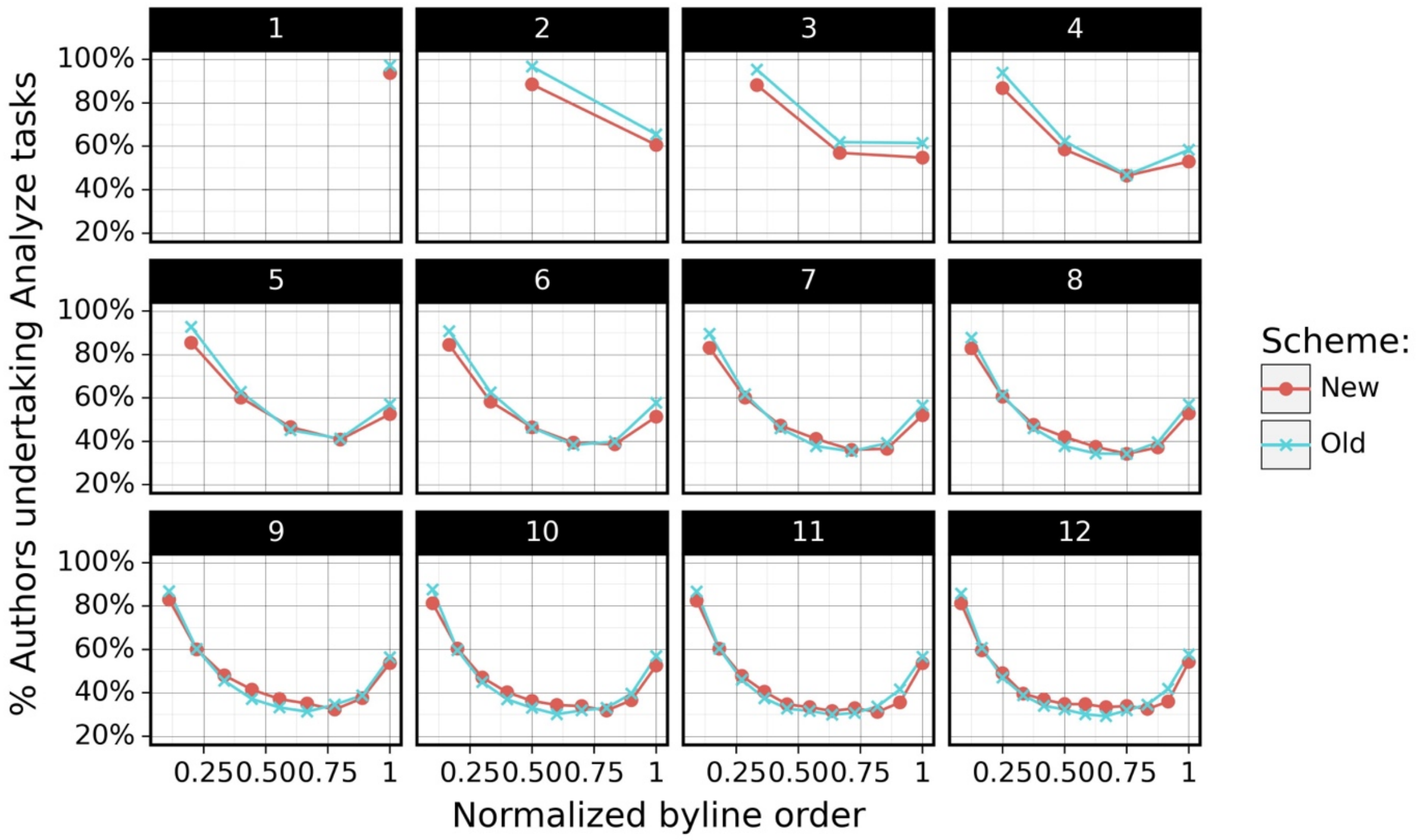

**Figure 9: Share of authors undertaking *Analyze* tasks by normalized byline order**

The above pattern is starkly contrasted to the *Write* category, an outlier in the authorship analysis. As shown in Figure 10, this category of tasks is much more likely to be performed by middle authors in the new system than in its precedent, which is contributed by *Writing review* as discussed above. Similarly, the *Perform* category also shows similar yet different curves between the two systems (as shown in Figure A1).

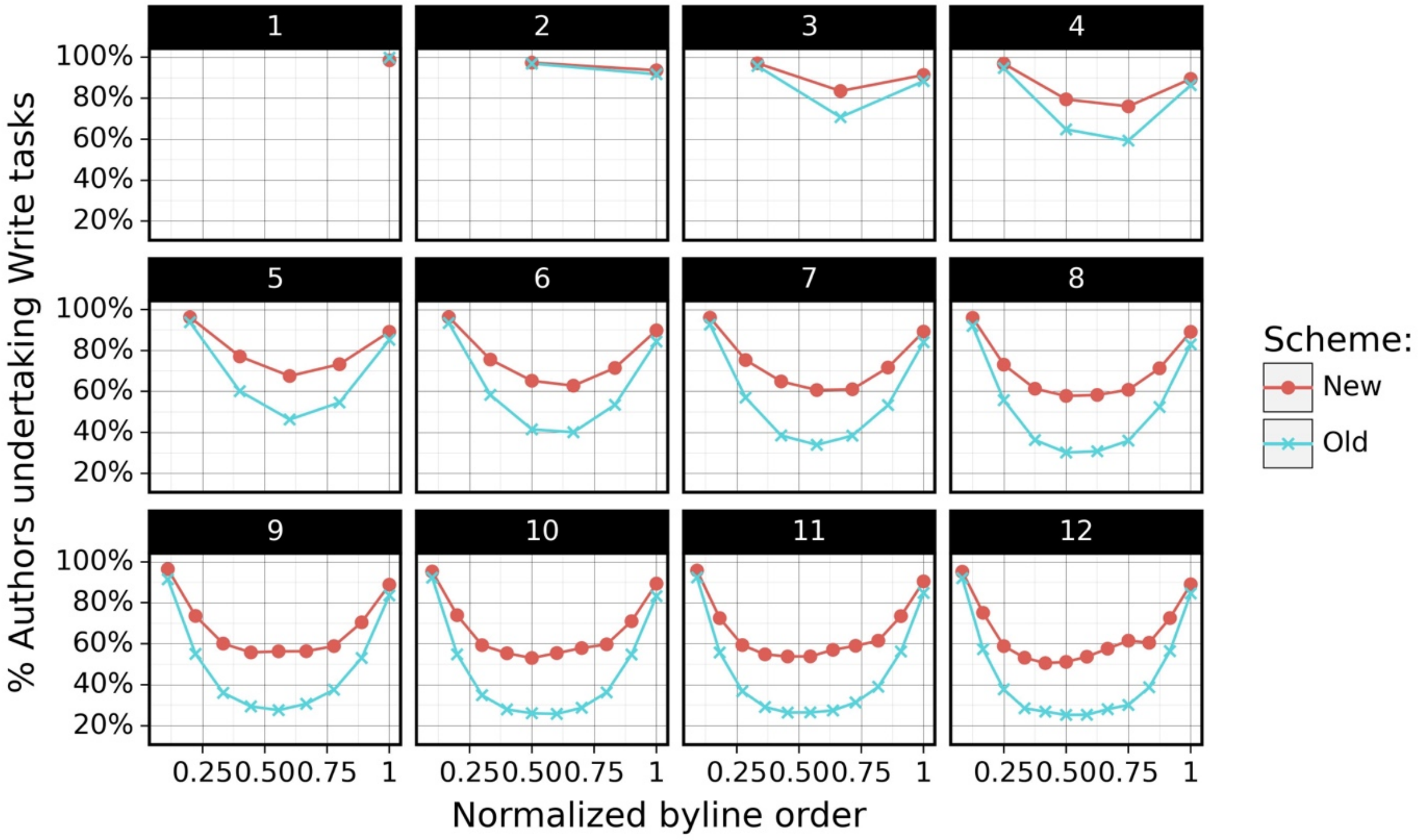

**Figure 10: Share of authors undertaking *Write* tasks by normalized byline order**

Another distinct category is *Contribute*. As shown in Figure 11, the new tasks in this category are more likely to be performed by first and last authors than the old tasks. In addition to the fact that not all the new tasks can be safely mapped to this category, this finding also indicates that this more technical-oriented category is more strongly highlighted in the CRediT system, being consistent with its designing goals.

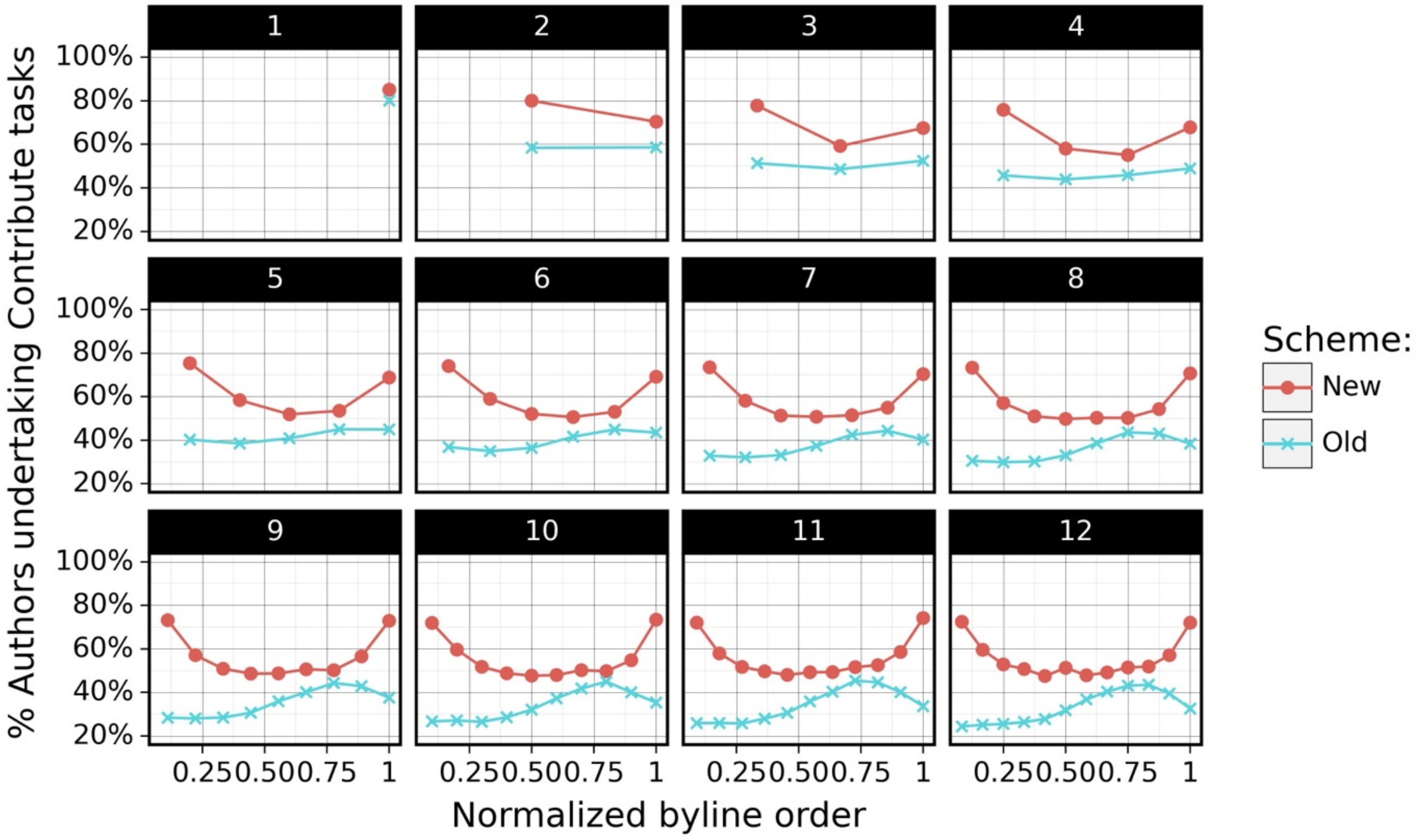

**Figure 11: Share of authors undertaking *Contribute* tasks by normalized byline order**

To better understand the individual tasks classified into *Contribute*, we further analyzed the pattern in Figure 11 on the level of task, as shown in Figure 12. There are a few notable observations. First, *Data curation* has a distinct pattern from the other new tasks, because it is much more likely to be assigned to top authors than all the other old and new tasks. It clearly shows that Data curation does not fit into the scope of *Contribute*. The same can be said about *Funding acquisition*, because it is much more likely to be undertaken by the last author, especially in papers with more authors. Second, for the other two tasks, combining them together still does not match with the old *Contribute* task. The above observation suggests that there are significant ontological differences between the old and new tasks around *Contribute*, which makes this category very difficult to be mapped between the two systems, a major challenge for future quantitative science studies.

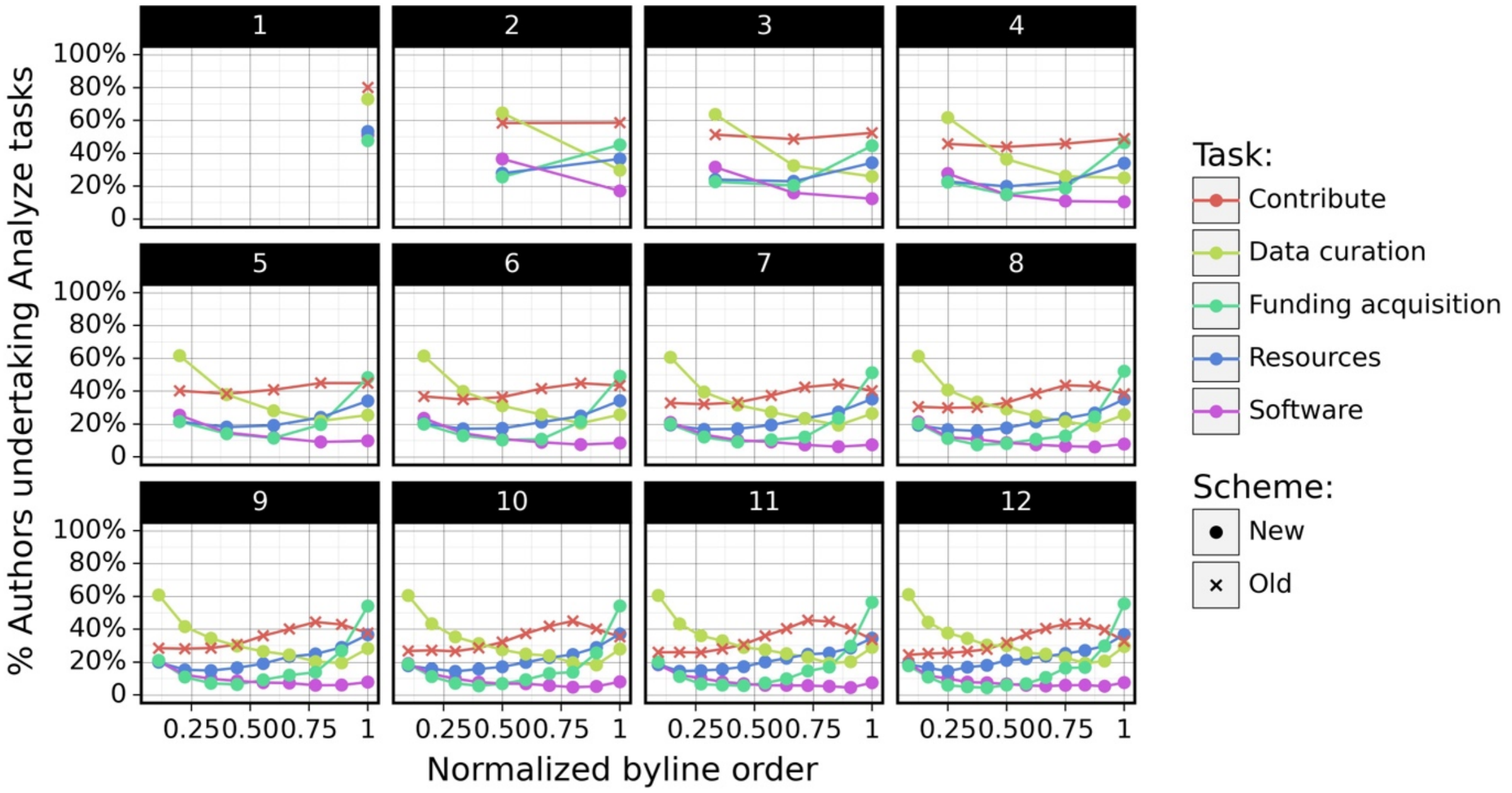

**Figure 12: Share of authors undertaking EACH *Contribute* task by normalized byline order (facet representing the different team sizes)**

The graphs for the rest of the two categories, i.e., *Perform* and *Conceive and design*, are offered in the Appendix. For all the patterns in these graphs, similar results are found if we choose a shorter publication window before and after 2016.

### 4.4 Distribution of tasks over researcher genders

This section focuses on the gender distribution of researchers undertaking each task in both classification systems. Table 3 summarizes the share of women authorship in each task. In most tasks, most researchers are male, echoing the disadvantage of women researchers on the level of authorship (Filardo et al., 2016; Macaluso et al., 2016). Moreover, in the tasks of *Perform* and *Investigation*, the two categories that are focused on "doing research," there is the most even distribution of researcher between the two genders. As compared, women researchers are less likely to contribute to categories that are more technical (such as *Software*) or require a higher social status (such as *Funding acquisition* and *Supervision*).

**Table 3: Share of women authorship undertaking each task**

| Task | Scheme | Authorship (men or women) | Share of women authorship |
|---|---|---|---|
| Perform | Old | 379,693 | 43.5% |
| Investigation | New | 220,589 | 43.1% |
| Data curation | New | 168,169 | 41.9% |
| Writing original | New | 162,469 | 41.2% |
| Formal analysis | New | 183,275 | 40.9% |
| Visualization | New | 88,379 | 40.1% |
| Methodology | New | 224,146 | 40.0% |
| Writing review | New | 315,601 | 38.5% |
| Validation | New | 112,018 | 37.8% |
| Project admin | New | 102,797 | 37.8% |
| Analyze | Old | 385,382 | 37.5% |
| Conceptualization | New | 211,798 | 37.0% |
| Write | Old | 418,437 | 35.7% |
| Resources | New | 112,356 | 34.2% |
| Conceive and design | Old | 359,851 | 34.1% |
| Funding acquisition | New | 92,165 | 33.6% |
| Contribute | Old | 286,084 | 33.1% |
| Supervision | New | 137,513 | 32.5% |
| Software | New | 54,359 | 30.9% |
| **All tasks** | | 4,015,081 | 37.9% |

Based on the five-year windows before and after 2016 for each classification system, we calculated the total share of women authorship in the five major categories based on our mapping scheme. The results are shown in Figure 13, which shows that in most cases, the difference between the two systems is relatively small, with the old system having fewer women researchers.

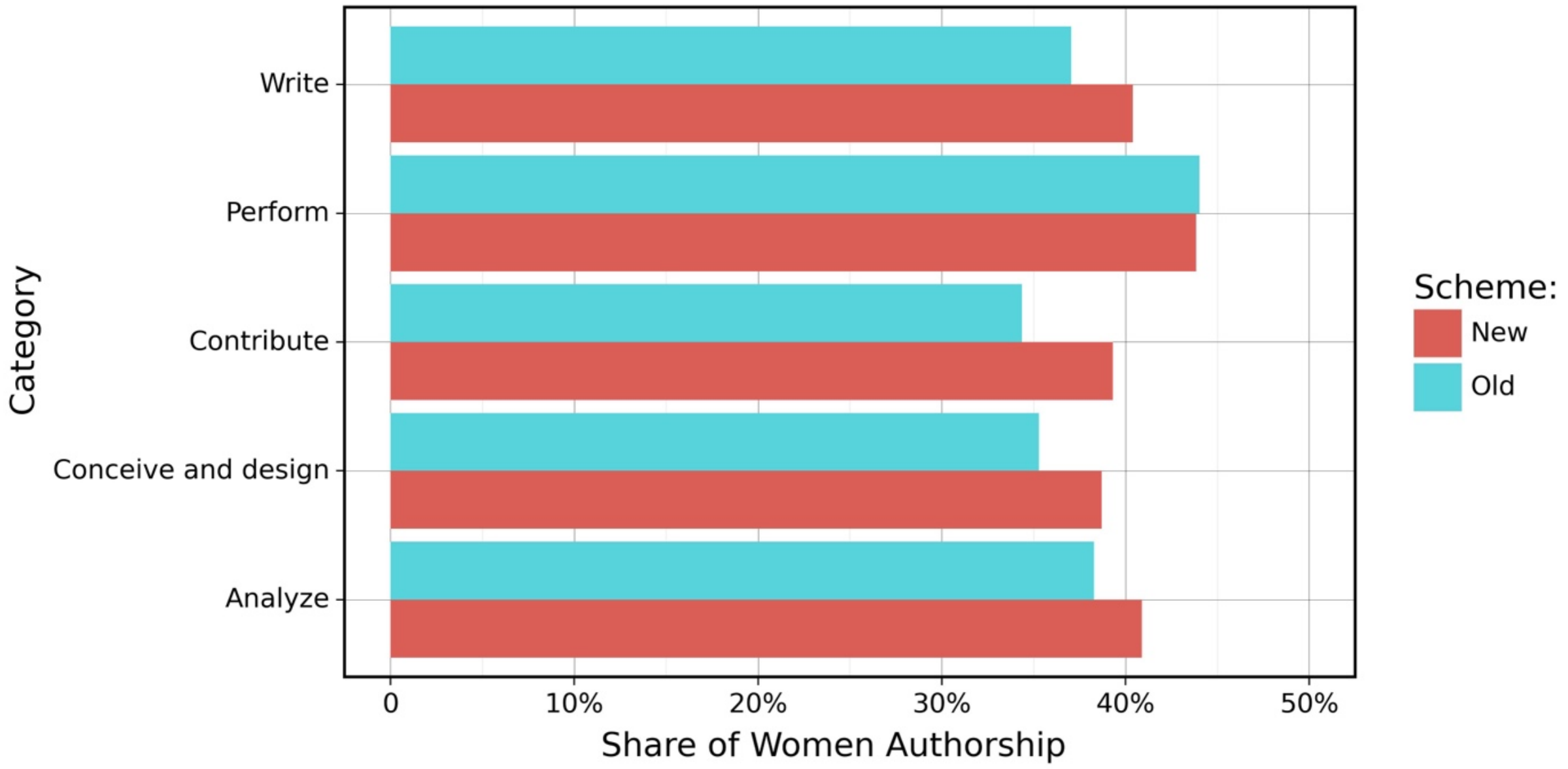

**Figure 13: Share of women authorship in each category of tasks**

The above pattern is further analyzed from the perspective of publication year. As shown in Figure 14, there is a very small gap in the year of 2016 in most of the categories and the old and new systems in each category form a linear, rising trend over time for the share of women authors. However, the only outlier to this overall pattern is *Perform*, where there is a mild gap during the transition of classification and the percentage of women researchers has not increased during the publication window.

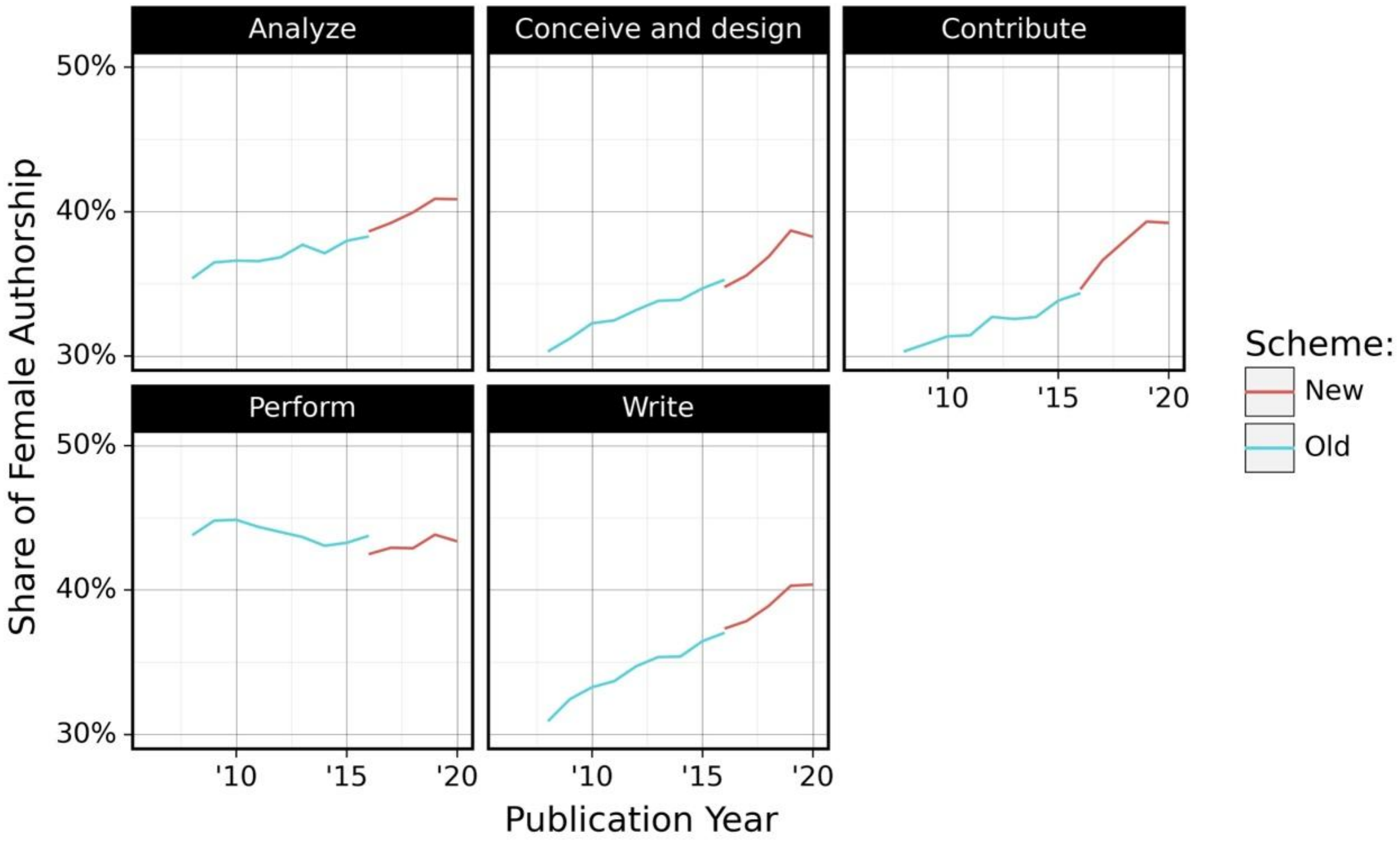

**Figure 14: Share of women authorship by task category over time**

## 5 Discussion

This study is the first effort to map the different contribution classification systems, using the example of PLoS ONE, to understand how researchers use the tasks in different classification systems. It bridges an important gap in empirical studies on division to labor due to the incompatibilities between the two classification systems used to assign research tasks to researchers. To better understand how the two systems are used, we designed a mapping scheme, based on which we further evaluated how the two systems are used by researchers in PLoS ONE before and after the year of 2016 from the perspectives of publications, authorship, byline order, and gender.

Our results suggest that even though the two classification systems have notable inconsistencies, they can largely be mapped to each other. Our mapping scheme covers all the five major tasks in the old classification system and 11 of the 14 new tasks in the CRediT system. The three other categories that cannot be mapped to the old system include *Funding acquisition*, *Project administration*, and *Supervision*. According to Holcombe (2019), one distinction between the contributorship model and the authorship model is that the former is not writing-mandatory and hence is broader than how authorship is defined by international policies, such as the guideline formulated by the International Committee of Medical Journal Editors (ICMJE). The above tasks can be understood from this perspective and all of these new tasks are used frequently on the publication-level, as shown in Table 2. More recent research has also aimed to further expand the CRediT model to include more contributions in research projects (Vasilevsky et al., 2021). The fluidity of the contributorship model makes it more challenging to create a stable dataset to be used by quantitative researchers.

Despite the incompatibilities and challenges discussed above, our analyses also show that our mapping scheme could work in most of the five categories due to the following reasons. First, many of the old and new tasks are naturally corresponding to each other, especially those in the *Write* and *Conceive and design* categories, which share very similar definitions. And second, the use of many categories shows strong consistencies in our multi-facetted comparison.

More specifically, categories of *Analyze*, *Conceive and design*, and *Perform* are used the most similarly across the two classification systems. Their usage patterns are generally highly consistent across the analyses we conducted, with only a few exceptions. Among them, *Conceive and design* and *Perform* are the only two one-on-one matches in our scheme, which can be more directly used to understand how the tasks in the two systems are used. For both of these categories, a notable observation is that they (along with *Analyze*) are performed by a smaller share of authors in the new classification system compared to the old system. This can be explained by the fact that given the larger number of tasks in the system, the threshold of task assignment is also higher (though the difference is small) in the new system.

The other two categories, however, are relatively less regular in how the tasks are used. For the category of *Write*, its mapping is supposed to be very reliable, and the category is used in the majority of publications in both systems, given the centrality of writing-mandatory in the definition of authorship. However, we find that *Writing review* is much more likely to be assigned to middle authors than other tasks in the same category and the difference is even larger for publications with larger teams. This could suggest that, under the writing-mandatory requirement, *Writing review* could be used as a sort of compensatory task to those who are not playing important roles in the research. And this could be an interesting research topic for a future study.

Lastly, we find that there are major differences in the category of *Contribute*, that no single or combination of new tasks could perform similarly with the old *Contribute reagents/materials/tools* task. In particular, *Data curation* and *Funding acquisition* are more likely to be performed by first and last authors, respectively, than the old task, and there are fewer authors performing *Resource* and *Software* tasks combined than the old task. The results indicate that this is the only category of tasks that may not be translatable between the two systems, given the very stark ontological differences between the tasks.

## 6 Conclusion

In the present paper, we proposed a mapping scheme between the two contributorship classification systems adopted by PLoS journals and used all publications from PLoS ONE to evaluate the scheme as well as how the two classification systems were used by researchers. The valuation is based on multiple perspectives, ranging from publications, authorship, byline order and gender, for a more comprehensive understanding of how researchers made use of the different classification systems and whether we can map the two systems to acquire a larger sample to investigate division of labor in scientific research. Our results show a strong possibility to map most of the categories in the two classification systems, which can be used to support future studies. Moreover, our results also indicate that as the new classification system hosts more categories, there is a generally higher threshold for researchers to be assigned a task and some tasks, such as *Writing review*, have distinct patterns from others in the same category.

Our paper makes timely contributions to studies on contributorship classification systems. It offers fresh empirical evidence on how researchers use different classification systems to assign contributions among all collaborators, by comparing the usage of two consecutive systems adopted by the same journal. Furthermore, our results will support future efforts to map contributorship data based on these two classification systems in PLoS journals, if not between other classification systems in other journals. This is an important step towards larger-scale quantitative analysis on the emerging topic of using contributorship statements to investigate how research works are divided between collaborators.

Despite our contributions, we acknowledge that this paper has a few limitations. First, this research is based on a single journal, which limits the applicability of results from our study. Despite the fact that PLoS journals and particularly PLoS ONE are one of the most frequently used data source for this line of research, we aim to expand our research in the future to other journals that require contributorship statements create a larger dataset of contribution statements. Second, one perspective to understand contribution that is missing from this research is that from unique authors. We also plan to investigate how individual authors and research teams undertake different tasks from the two systems in the future, to better understand our research problem.

## Appendix

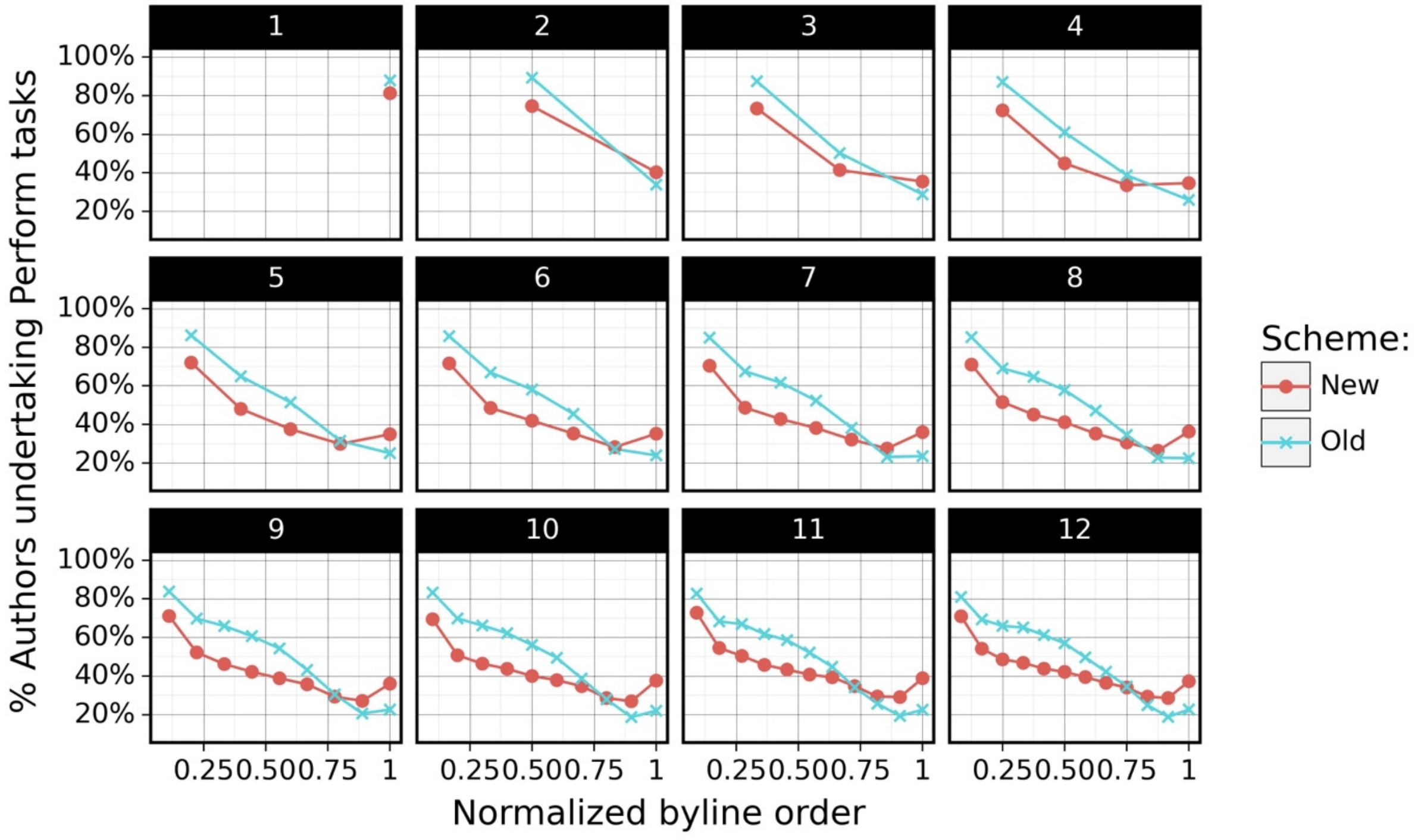

**Figure A1: Share of authors undertaking *Perform* tasks by normalized byline order**

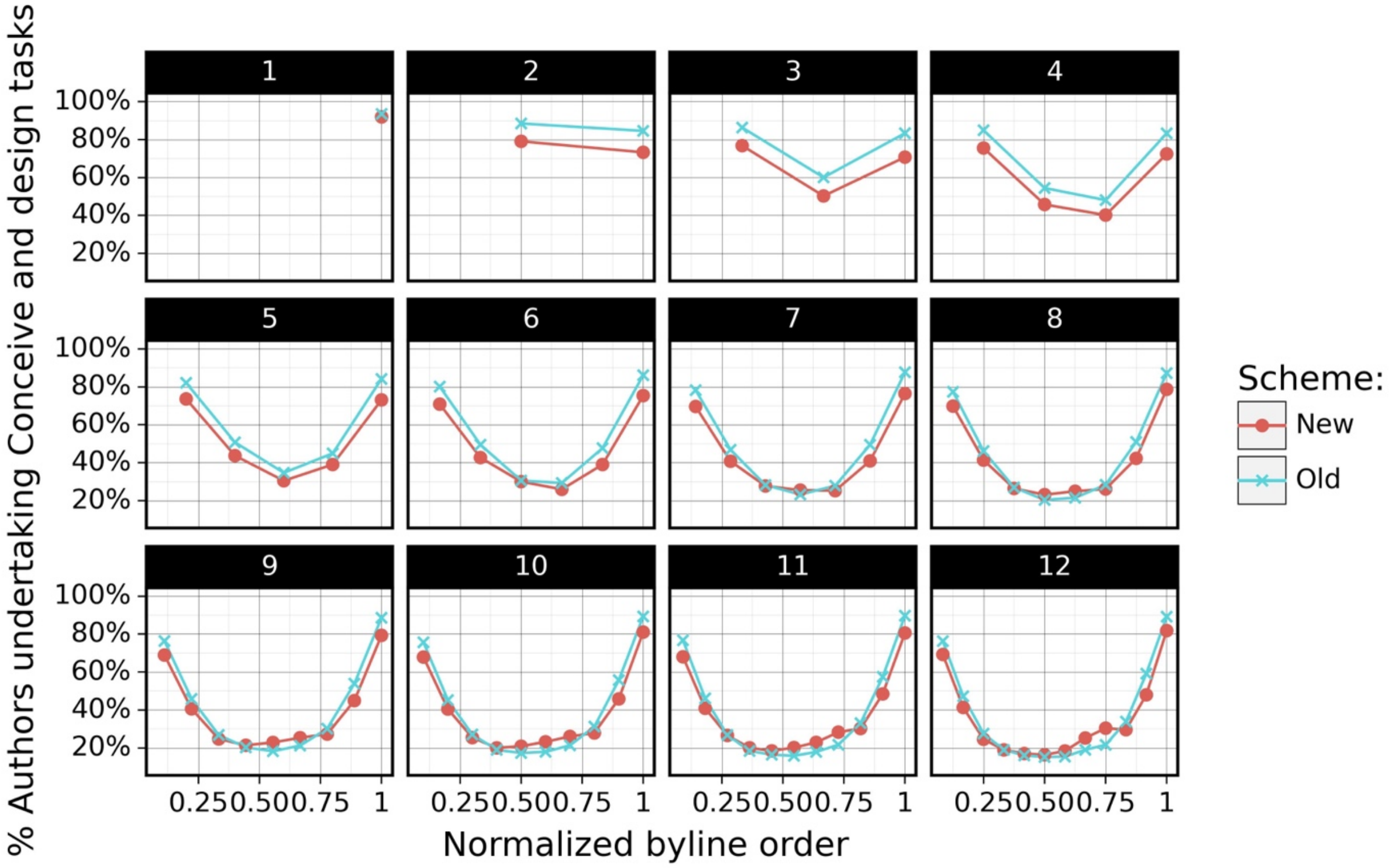

**Figure A2: Share of authors undertaking *Conceive and design* tasks by normalized byline order**